\documentclass{article}
\usepackage{amsthm,amsmath,amsfonts,amssymb}
\usepackage{natbib}
\usepackage{amsmath}
\usepackage{graphicx}
\usepackage{url} 
\usepackage[T1]{fontenc}
\usepackage[utf8]{inputenc}
\usepackage{float}
\usepackage{color,soul}
\usepackage{multicol} 
\usepackage{multirow}
\usepackage{booktabs}
\usepackage{ulem}

\newcommand{\re}{\mathbb R}

\begin{document}

\title{A Python implementation of some geometric tools on Kendall 3D shape space for practical applications}

\author{Jorge Valero$^{(1)}$, Vicent Gimeno i Garcia$^{(2)}$,  \\
        M. Victoria Ib\'a\~nez$^{(2)}$, Pau Martinavarro $^{(2)}$ and 
        Amelia Sim\'o$^{(2)}$\\
\small{(1) Instituto de Biomecánica de Valencia (IBV), Universitat Politècnica de València, Valencia, Spain}\\
\small{(2) Department of Mathematics-IMAC, Universitat Jaume I,
              Castell\'o, Spain}
}

\maketitle

\begin{abstract}
This work addresses the challenge of analyzing geometric structures using Kendall’s Shape Space.
While Riemannian geometry provides a robust framework for shape analysis (independent of scale, position, and orientation) the transition from theoretical manifolds to practical computational workflows remains difficult.
Although Geomstats is currently the leading Python library for manifold-based statistics, it lacks specific utilities required for advanced 3D shape analysis. 
This article introduces tools designed to bridge this gap, translating complex mathematical abstractions into efficient, accessible software solutions for researchers.

\textbf{Keywords}
 Kendall’s 3D Shape Space, Riemannian Geometry, Geomstats, Sectional Curvature, Tangent Orthonormal Basis
\end{abstract}

\section{Introduction}
In the field of computer vision and medical imaging, the analysis of geometric structures—independent of their position, scale, and orientation—is a fundamental challenge. Kendall’s Shape Space provides a rigorous mathematical framework for this task, treating shapes as points on a manifold and enabling the use of Riemannian geometry to perform statistical analysis. While the theoretical foundations of shape manifolds are well-established, translating these complex abstractions into efficient, practical computational tools remains a significant hurdle for many researchers.

Currently, Geomstats \citep{miolane2020geomstats} stands as the most prominent Python library for Riemannian geometry, offering a wide array of tools for manifold-based statistics. However, when it comes to the specific requirements of 3D shape analysis, practitioners often encounter notable missing utilities. 
This article aims to fill some gaps left by general-purpose libraries like Geomstats. 

\subsection{Mathematical Preliminaries: Kendall’s 3D Shape Space}
The shape space  $\Sigma^k_3$ is the set of equivalence classes of $k \times 3$ configuration matrices $X\in M_{k,3}(\mathbb{R})$ under the action of Euclidean similarity transformations. 

We can obtain a representative of each equivalence class by eliminating each transformation one by one. 
One way to remove the location effect consists of multiplying the configuration matrix, $X$, by the Helmert submatrix $H$ (\cite{DrydenMardia16}), obtaining the Helmertized landmark coordinates $X_H=(XH)^t$. Thus, $X_H \in M_{(k-1),3}(\mathbb{R})$.
To filter scale, $X_H$ can be divided by the centroid size, which is given by its Frobenius norm.
$Z_X=\frac{X_H}{\|X_H\|}$ is called the pre-shape of the configuration matrix $X$ because all information about location and scale is removed, but rotation information remains.

The pre-shape space,  $S^k_3$, is  the set of all possible pre-shapes. Thus, each pre-shape 
is a $ (k-1) \times 3 $ matrix with size one,  and as a result, it lies on  the 
hypersphere of unit radius in $\re^{3(k-1)}$. 

The shape space, $\Sigma^k_3$, is the quotient space of $S^k_3$ under rotations. A shape is an orbit associated with the action of the rotation group $SO(3)$ on the hypersphere of unit radius in $\re^{3(k-1)}$.  

The geometry of the shape space is intrinsically non-Euclidean. Given a point $X \in \Sigma^k_3$, the Tangent Space $T_X\Sigma^k_3$ provides a local linear approximation of the manifold. Using the Logarithmic Map, we can "project" shapes from the curved manifold onto this flat vector space. 

A critical feature of Kendall’s 3D shape space is its Sectional Curvature. Unlike simpler manifolds, the shape space  possesses non constant positive sectional curvature.
The sectional curvature $K(\sigma)$ measures how much the manifold deviates from Euclidean flatness along a 2-dimensional plane $\sigma$ spanned by tangent vectors $\{u,v\}$. 
It can be expressed:
$$ K(u,v)=\frac{g(R(u,v)v,u)}{g(u,u)g(v,v)-g(u,v)^2}$$
with $g$ the Riemann metric and $R$ the curvature tensor \citep{do1992riemannian}. 
 
\section{Orthonormal basis for simulation on the tangent space}

Some practical applications need to simulate on the tangent space at a point $X$ of the shape space. In these applications the representative of this point is on the hypersphere of unit radius in $\re^{3(k-1)}$ and it is expressed on the $3(k-1)$ embedded Euclidean coordinates. However, the dimension of the shape space and then of the Tangent space is $d=3k-7$.

This section is devoted to explain how to obtain an orthonormal basis $\{v_1, \ldots, v_{d}\}$ at any point $X$ in Kendall's shape space, which is required to perform the simulations. Referenced to this specific basis we can obtain the $d$ coordinates of any tangent vector.   

Our goal is to recast this problem as one of computing the Singular Value Decomposition (SVD) of a specific matrix, in order to leverage efficient and well-established numerical algorithms.

As it has been stated above, for any non-singular point $[Z] \in \Sigma^k_3$, $Z \in S^k_3$, and we can consider $S^k_3\subset \mathbb{R}^{3(k-1)}$. The theory of Riemannian submersions tell us that The tangent space of the shape space is the horizontal tangent space of the pre-shape sphere.

The tangent space $T_Z\mathbb{R}^{3(k-1)}\simeq \mathbb{R}^{3(k-1)}$ can be decomposed orthogonally as
\begin{equation}\label{equ:ortdesc}
\mathbb{R}^{3(k-1)}={\rm span}_{\mathbb{R}}(Z)\oplus\mathcal{V}_Z\oplus\mathcal{H}_Z,    
\end{equation}
We will use this fact to obtain an orthonormal basis of $\mathcal{H}_Z=T_X\Sigma^k_3$. 

Taking into account that the vertical distribution at $Z$  is given by:
$
\mathcal{V}(Z)=\left\{ZA\,:\, A^T=-A\right\}={\rm span}_{\mathbb{R}}\left\{{Z\boldsymbol {L}}_{x},\,  {Z\boldsymbol {L}}_{y},\,{Z\boldsymbol {L}}_{z} \right\}
$
with $\boldsymbol {L}_{x}$, $\boldsymbol {L}_{y}$ and $\boldsymbol {L}_{z}$ being the following basis for the Lie algebra $SO(3)$:
\begin{eqnarray}\label{base L}
\displaystyle {\boldsymbol {L}}_{x}={\begin{bmatrix}0&0&0\\0&0&-1\\0&1&0\end{bmatrix}},\quad {\boldsymbol {L}}_{y}={\begin{bmatrix}0&0&1\\0&0&0\\-1&0&0\end{bmatrix}},\quad {\boldsymbol {L}}_{z}={\begin{bmatrix}0&-1&0\\1&0&0\\0&0&0\end{bmatrix}},
\end{eqnarray}
and by using the orthonormal decomposition \eqref{equ:ortdesc} we can conclude that the space $\mathcal{H}_Z$ can be obtained as the kernel of the following endomorphism:

\begin{eqnarray}\label{Eq: Def_f}
A &\mapsto& f(A) := \langle A, Z \rangle Z 
+ \langle A, Z\boldsymbol{L}_x  \rangle Z\boldsymbol{L}_x 
+ \langle A, Z\boldsymbol{L}_y  \rangle Z\boldsymbol{L}_y 
+ \langle A, Z\boldsymbol{L}_z  \rangle Z\boldsymbol{L}_z. 
\end{eqnarray}

Thence, to obtain a orthonormal basis for $\mathcal{H}(Z)$ is equivalent to obtain an orthonormal basis for the kernel of $f$. On the other hand, Since $f$ is a self-adjoint linear operator, there exists $P\in SO(3(k-1))$ such that 
\[
M = P \Lambda P^T,
\]
where $M$ is the matrix representation of $f$ with respect to the canonical basis of $\mathbb{R}^{3(k-1)}$, and $\Lambda$ is a diagonal matrix.

We then construct an orthonormal basis of $\mathcal{H}_Z$ by selecting the column vectors $\{t_1, \ldots, t_{d-4}\}$ of $P$ corresponding to the zero eigenvalues in $\Lambda$. Any tangent vector $V \in \mathcal{H}_Z$ can thus be expressed in coordinates as $V=(v_1,v_2,\cdots, v_{d-4})$ since
$V = \sum_{i=1}^{d-4} v_i t_i$, where   $v_i = \langle V, t_i \rangle$.

\section{Sectional Curvature Calculation}
Kendall’s formula for sectional curvature in the shape space is notably more complex than in a standard sphere because of the quotient map.

For two orthonormal tangent vectors $u, v \in T_X\Sigma^k_3$, the sectional curvature $K(u,v)$ is given by:
\begin{eqnarray}
K(u,v)=1+\frac{3}{4}\left\Vert[u,v]^\mathcal{V}\right\Vert^2,
\end{eqnarray}\label{Eq: calculo curvatura}
where $[u,v]^\mathcal{V}$ represents the vertical component of the bracket. 

The primary challenge in this expression is the computation of the bracket’s vertical components, because there are not an known expression for any $v$  in $T_X\Sigma^k_3$. But fortunately only the module is needed and in page 148 of \cite{Kendalletal09} we can find the important products for the vectors of a given basis of $T_X\Sigma^k_3$. Then, our procedure is based on calculate the vector of this basis, make a change of basis and finally to use these known products. 

\subsection{Deriving the Required Basis for the Tangent Space at $X$}\label{Sec: Base}

The following steps outline the procedure to obtain the tangent space basis necessary for calculating the sectional curvature:

\begin{itemize}
    \item[\textbf{Step 1:}] \textbf{Singular Value Decomposition (SVD):} Perform the decomposition of the point $X$ where the sectional curvature is to be evaluated: $X = U \Sigma V^T$. Hence,
$$
\Sigma=(\Lambda,0)=\left(\begin{pmatrix}
\lambda_{1} & 0&   0  \\
 0 & \lambda_{2} &  0  \\
 0 & 0 & \lambda_{3} 
    \end{pmatrix} ,\begin{pmatrix}
     0 & \cdots & 0\\
      0 & \cdots &0\\
       0 & \cdots &0\\
\end{pmatrix}\right)
$$
    
\item[\textbf{Step 2:}] \textbf{Obtaining $Z_{ij}$:} 
    Following Equation (7.2) on page 137 of \cite{Kendalletal09}, $Z_{ij}$ is defined as $Z_{ij} = U W_{ij} V$, where:
    \begin{itemize} 
        \item[] $W_{ii}=E_{ii}-\frac{\lambda_i}{\lambda_1}E_{11}$, for $1 < i \leq 3$.
        \item[] $W_{ij}=\Lambda (E_{ij}+E_{ji})$, $1 \leq i < j \leq 3$.
        \item[] $W_{ij}=\Lambda E_{ij}$, $1 \leq i \leq m < j \leq k-1$
    \end{itemize}

    \item[\textbf{Step 3:}] \textbf{Basis Transformation:} 
    Following Corollary 7.1 on page 146, we construct a new basis $\{\frac{\partial}{\partial \lambda_i}\}\cup\{\xi_{ij}\}$. We identify $Z$ with the vector composed of the matrix elements of $Z$ and its corresponding projection:
    \begin{itemize}
        \item[] $\frac{\partial}{\partial \lambda_i}=Z_{ii}$,\quad  $1 < i \leq 3$.
        \item[]  $ \xi_{ij}=\frac{ \left(\lambda^2_i-\lambda^2_j\right)}{\lambda^2_i+\lambda_j^2}Z_{ij}$, \quad $1 \leq i < j \leq 3$.
        \item[] $\xi_{ij}=\lambda_i Z_{ij}$,\quad $1 \leq i \leq 3 < j \leq k-1$.
    \end{itemize}
\end{itemize}

\subsection{Change of Basis and Curvature Calculation}\label{Sec:Calculo curvatura}

\begin{itemize}
    \item \textbf{Representation:} Express the vectors $u$ and $v$ in the Kendall basis.
    
    \item \textbf{Bilinearity:} Applying the bilinearity of the Lie bracket operator ($[u,v]^\mathcal{V}$), we utilize the relevant commutators found between the end of page 151 and the beginning of page 152:

$$
\begin{aligned}
    \left[\frac{\partial}{\partial \lambda_i}, \frac{\partial}{\partial \lambda_j}\right]^\mathcal{V} &= 0 \\[15pt]
    \left[\frac{\partial}{\partial \lambda_l}, \xi_{ij}\right]^\mathcal{V} &=2 
    \begin{cases} 
        \displaystyle \frac{1}{\lambda_1} \frac{\lambda_1^2 - \lambda_j^2}{\lambda_1^2 + \lambda_j^2} \eta_{ij} & i = 1 \text{ and }1< l = j\leq 3, \\[10pt]
        \displaystyle \frac{\lambda_j \lambda_l}{\lambda_1} \frac{\lambda_1^2 - \lambda_j^2}{(\lambda_1^2 + \lambda_j^2)^2} \eta_{ij} &  i = 1 \text{ and }1< l \neq j \leq 3, \\[10pt]
        \displaystyle \lambda_j \frac{\lambda_j^2 - \lambda_i^2}{(\lambda_i^2 + \lambda_j^2)^2} \eta_{ij} &  1 < i \leq 3 \text{ and } l = i<j\leq 3, \\[10pt]
        \displaystyle \lambda_i \frac{\lambda_i^2 - \lambda_j^2}{(\lambda_i^2 + \lambda_j^2)^2}\eta_{ij} &  1 < i \leq 3 \text{ and } i< l = j \leq 3, \\[10pt]
        0 & \text{otherwise.}
        \end{cases} \\
[\xi_{ij_1}, \xi_{ij_2}]^\mathcal{V} &= -[\xi_{ij_1}, \xi_{j_2i}]^\mathcal{V} \\
&= \left\{ -4 \frac{\lambda_i \lambda_{j_1}}{\lambda_i^2 + \lambda_{j_1}^2} \frac{\lambda_i \lambda_{j_2}}{\lambda_i^2 + \lambda_{j_2}^2} + 2 \frac{\lambda_{j_1} \lambda_{j_2}}{\lambda_{j_1}^2 + \lambda_{j_2}^2} \right\} \eta_{j_1j_2}, \\
&\quad 1 \leq i < j_1 \leq m \text{ and } 1 \leq i \neq j_2 \leq m, \\
[\xi_{i_1j}, \xi_{i_2j}]^\mathcal{V} &= -[\xi_{i_1j}, \xi_{ji_2}]^\mathcal{V} \\
&= \left\{ -4 \frac{\lambda_{i_1} \lambda_j}{\lambda_{i_1}^2 + \lambda_j^2} \frac{\lambda_{i_2} \lambda_j}{\lambda_{i_2}^2 + \lambda_j^2} + 2 \frac{\lambda_{i_1} \lambda_{i_2}}{\lambda_{i_1}^2 + \lambda_{i_2}^2} \right\} \eta_{i_1i_2}, \\
&\quad 1 \leq i_1 < j \leq 3 \text{ and } 1 \leq i_2 \neq j \leq 3, \\
[\xi_{i_1j}, \xi_{i_2j}]^\mathcal{V} &= 2 \frac{\lambda_{i_1} \lambda_{i_2}}{\lambda_{i_1}^2 + \lambda_{i_2}^2} \eta_{i_1i_2}, \quad 1 \leq i_1, i_2 \leq 3 < j \leq k - 1.
\end{aligned}
$$

\item \textbf{Inner Products:} The essential inner products for $\eta_{ij}$ are provided in Equation (7.12) on page 148.
$$
\langle \eta_{i_1j_1},\eta_{i_2j_2}\rangle=\delta_{i_1i_2}\delta_{j_1j_2}\left(\lambda_{i_1}^2+\lambda_{i_2}^2\right).
$$
    \item \textbf{Final Curvature Calculation:} By substituting the calculated commutators into the expression for the sectional curvature and applying the inner products from Equation (7.12), the vertical components are fully determined. This allows for the final evaluation of the curvature at $X$, effectively combining the geometric information of the Kendall basis with the specific configuration of the shape space.
\end{itemize}

\section{Python Implementations}

We extend the current Geomstats functionality by adding the following functions:

\subsection{For the orthonormal basis for simulation}
File \textbf{orthonormal\_basis\_to\_simulate.py} is structured in the following functions:
\begin{itemize}
    \item \textbf{get\_tangent\_base\_kendall}: that computes the orthonormal basis of the tangent space to Kendall's shape space at a given point.
    \item \textbf{simulate\_in\_tangent\_space}: uses the previous function to generate random samples in the tangent space and project them onto the shape space.
\end{itemize}

A final example is provided to simulate 3D houses defined by 10 landmarks, in the tangent space to the shape space at a given house (casa\_ref).

\subsection{For the sectional curvature calculation}

File \textbf{sectional\_curvature.py} is structured in the following functions:

\begin{itemize}
\item \textbf{\_init\_}: Performs the Singular Value Decomposition (SVD) of the input configuration matrix $X$ to extract the matrix $U$, singular values $\lambda_i$, and the matrix $V$. (Step 1. Sec. \ref{Sec: Base} )

\item \textbf{get\_xi\_basis}: Constructs the adapted orthonormal basis $\{\xi_{ij}\}$ for the horizontal tangent space. (Step 2 and 3. Sec. \ref{Sec: Base} )
\item \textbf{compute\_vertical\_bracket\_norm\_sq}: Calculates the squared norm of the vertical component of the Lie bracket, $\lVert [u, v]^\mathcal{V} \rVert^2$, by applying the specific bilinear commutator formulas and $\eta_{ij}$ products. (Sec. \ref{Sec:Calculo curvatura} )
\item \textbf{sectional\_curvature}: Computes the sectional curvature for a plane defined by two vectors $u$ and $v$ using the Eq. \ref{Eq: calculo curvatura}: $sec(u \wedge v) = 1 + \frac{3}{4}\lVert [u, v]^\mathcal{V} \rVert^2$. 
\end{itemize}

A final function called \textbf{compute\_curvature} allows to execute the curvature calculation for a specific example.

The software can be found in \\
\url{https://sites.google.com/uji.es/toia/research}


\end{document}